\documentclass[graybox]{svmult}

\usepackage{mathptmx}       
\usepackage{helvet}         
\usepackage{courier}        
\usepackage{type1cm}        
\usepackage{makeidx}         
\usepackage{graphicx}        
\usepackage{multicol}        
\usepackage[bottom]{footmisc}

\makeindex             

\begin{document}

\title*{Testing models of triggered star formation: theory and observation}
\author{Thomas J. Haworth, Tim J. Harries and David M. Acreman}
\institute{Thomas Haworth \at University of Exeter, \email{haworth@astro.ex.ac.uk}}
%
%
\maketitle

\abstract*{}

\abstract{One of the main reasons that triggered star formation is contentious is the  failure to accurately link the observations with models in a detailed, quantitative, way. It is therefore critical to continuously test and improve the model details and methods with which comparisons to observations are made. We use a Monte Carlo radiation transport and hydrodynamics code \textsc{torus} to show that the diffuse radiation field has a significant impact on the outcome of radiatively driven implosion (RDI) models. We also calculate SEDs and synthetic images from the models to test observational diagnostics that are used to determine bright rimmed cloud conditions and search for signs of RDI.}
\vspace{40pt}

We have investigated the impact of polychromatic and diffuse field radiation on radiatively driven implosion (RDI) models using the Monte Carlo radiation transport and hydrodynamics code \textsc{torus} \cite{H00,2012MNRAS.420..562H}. The details of the code implementation, model parameters and results are given in \cite{2012MNRAS.420..562H}. We ran three types of RDI calculation. One with a monochromatic radiation field, one with a polychromatic radiation field and one that is both polychromatic and includes the diffuse radiation field. The addition of polychromatic radiation to the calculation does not significantly alter the outcome of the model. However, including the diffuse field can lead to significantly different evolution of the cloud, altering the morphology and increasing the maximum accumulated density after 200\,kyr up to about a factor of 10. 

\begin{figure}[t]
\sidecaption
\includegraphics[scale=.2]{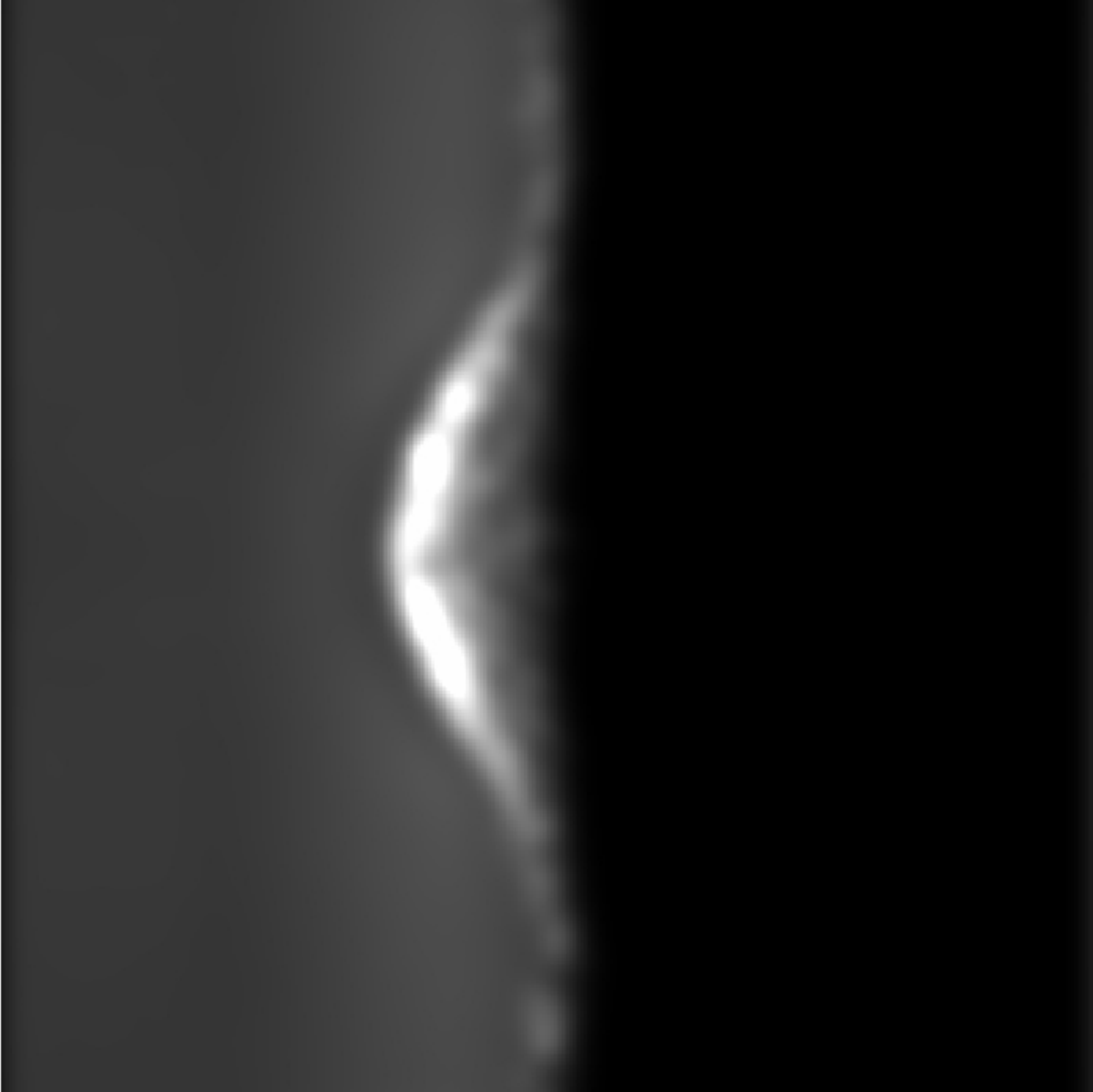}
%
%
\caption{A simulated 20\,cm continuum emission image of a BRC taken using the VLA configuration C. Synthetic images like this can be analyzed in the same way as real data, which is useful for testing the accuracy of diagnostics and improving the interpretation of characteristic BRC features.}
\label{fig:low_BRC_VLAC}       
\end{figure}

Using these RDI models from \cite{2012MNRAS.420..562H} we calculated synthetic images and SEDs to test observational diagnostics of bright rimmed clouds (BRCs) in \cite{HHD12}. We calculated the neutral cloud properties by fitting the cloud SED as a greybody to determine the dust temperature, which can then be used to calculate the cloud mass following \cite{H83}. The temperature and electron density in the ionized boundary layer (IBL) and the cloud mass loss rate were calculated using simulated VLA 20\,cm continuum images, an example of which is given in Figure \ref{fig:low_BRC_VLAC}, and the standard techniques of \cite{LL94, LL97}. We have also tested the use of forbidden line ratios from long slit spectroscopy to determine the IBL conditions and found that they are a viable tool, giving a direct and more accurate measure of the IBL temperatures compared to the radio method which assumes a canonical value of 10$^4$\,K.

Using the inferred cloud and IBL conditions we calculated the cloud support and IBL pressures to determine whether or not the clouds are being compressed. We find that this pressure comparison diagnostic is a reasonable indicator of whether or not the IBL is driving into the cloud. The accuracy of the techniques was investigated by comparing the derived conditions and behaviours with those known from the model grid. For example, we have demonstrated that as the beam size increases the IBL conditions are increasingly underestimated in the radio diagnostic because the IBL flux is contaminated by the neutral cloud and HII region. We also found that the contribution to the SED from warm dust causes a slight overestimation of the dominant cloud temperature by $1-2$\,K, which leads to an overestimation of the mass by up to a factor of 35\%. Furthermore, use of a constant mass conversion factor $C_{\nu}$ in the mass calculations of \cite{H83} for BRCs of different class is found to introduce errors up to a factor 3.6. This comparison of the known conditions in simulations with those inferred through observational diagnostics of synthetic data means that more reliable conclusions can be drawn from studies of real BRCs.

%

\bibliographystyle{spphys}
\bibliography{haworth}

\end{document}